\documentclass[a4paper]{jpconf}
\usepackage{graphicx}
\newcommand{\JJ}{J}
\newcommand{\pp}{p}
\newcommand{\qq}{q}

\newcommand{\Jsl}{\JJ\hspace{-7pt}/}
\newcommand{\ksl}{k\hspace{-6pt}/}
\newcommand{\esl}{e\hspace{-5.5pt}/}
\newcommand{\Ta}{T^{a}}
\newcommand{\kA}{k_1}
\newcommand{\eA}{e_1}
\newcommand{\ksA}{\ksl_1}
\newcommand{\esA}{\esl_1}
\newcommand{\Tb}{T^{b}}
\newcommand{\kB}{k_2}
\newcommand{\eB}{e_2}
\newcommand{\ksB}{\ksl_2}
\newcommand{\esB}{\esl_2}

\newcommand{\inp}[2]{#1\!\cdot\!#2}

\newcommand{\IAB}{I^{(1,2)}}
\newcommand{\IBA}{I^{(2,1)}}

\newcommand{\Mab}{\mathcal{M}^{a,b}}

\newcommand{\formula}[1]{\mbox{#1}}

\newcommand{\Meu}{{\cal M}}
\def\st{\hbox{}}

\begin{document}
\title{Automated calculation of matrix elements and physics motivated observables}

\author{Z. Was}

\address{Institute of Nuclear Physics, Polish Academy of Sciences 31 342 Krakow, Poland}

\ead{z.was@cern.ch}

\begin{abstract}
The central aspect of my personal scientific activity, has focused on calculations useful for interpretation of High Energy
accelerator experimental results, especially in a domain of precision tests of the Standard Model. 
My activities started in early 80's, when computer support for algebraic
manipulations was in its infancy. But  already then it was important for my work. It brought a multitude of benefits, but at 
the price of some inconvenience for physics intuition. Calculations became more complex, work had to be distributed 
over teams of researchers and due to automatization, some aspects of the intermediate results became more difficult to identify.

In my talk  I will not be very exhaustive, I will present  examples from my personal research only:  
(i) calculations of spin effects for the process  $e^+ e^- \to \tau^+\tau^- \gamma$ at Petra/PEP energies, 
 calculations (with the help of the {\tt Grace} system of Minami-tateya group) and phenomenology of spin amplitudes for 
(ii) $e^+ e^- \to 4 { f}$
 and for  (iii) $e^+ e^- \to \nu_e \bar \nu_e \gamma \gamma$ processes, (iv) phenomenology of CP-sensitive observables
for Higgs boson parity in $H \to \tau^+ \tau^- ,\; \tau^\pm \to \nu 2(3)\pi$ cascade decays.

\vskip 3 mm
\centerline{\bf  Presented at ``4th Computational Particle Physics Workshop'' 8 - 11 October 2016, Hayama Japan} 

\centerline{\bf Preprint IFJ PAN-IV-2017-2, January 2017}
\end{abstract}

\section{Introduction}

 Once computers equipped with algebraic languages became available, approaches to phenomenology of 
High Energy accelerator experiments changed substantially. For me personally, it all started in  1980.
It was clear that numerous benefits appeared. 
Control of large expression became easier, in many cases it simply became possible for the first time.
 Nearly immediately  
some drawbacks appeared as well.  For example, some seemingly obsolete expertise, like 
methods for special function expansions, started to disappear. At least for some years and for some communities.

This was all part of a complex and generally very fruitful development. In this  presentation
I will concentrate on my personal experience. I do not have any intentions to be systematic or 
balanced. A more balanced  picture will hopefully appear together with other talks collected in the proceedings.
That is also why, I think, I do not need to focus  on successes of the field. These are well known.
I will rather review difficulties or traps I have encountered myself. In fact traps, once resolved, turned out 
to be rewarding, often in an unexpected way. 

The presentation is organized as follows. Section \ref{sec:KORALB} will  discuss computer algebra techniques 
which were applied in our work for Monte Carlo program {\tt KORALB}  \cite{Jadach:1984iy} for the
  $e^+ e^- \to \tau^+\tau^- \gamma$ process at Petra/PEP energies. The following
Section \ref{sec:kkmc}, is devoted to work on spin amplitudes for {\tt KKMC} Monte Carlo \cite{Jadach:1999vf}. In particular 
to studies of spin amplitude sub-structure for the $e^+ e^- \to \nu_e \bar \nu_e \gamma \gamma$ process, necessary  to accommodate
resulting expressions to  Yennie Frautschi Suura exponentiation \cite{yfs:1961};
variant of
 exclusive exponentiation of initial and final state radiation \cite{Jadach:2000ir}. The t-channel (even though non-singular)  
$W$ exchange required careful attention.  The following Section \ref{sec:4f} is devoted to phenomenology of 
four-fermion state production at high energy and the study of unexpected formation of peaks due to conspiracy between 
spin and selection cut effects~\cite{Fujimoto:2002sj}.  Finally, in Section \ref{sec:ML},  we turn our attention to  methods of Machine Learning in 
applications for evaluation of (massively multi-dimensional) observables for Higgs parity~\cite{Jozefowicz:2016kvz}.
Section \ref{sec:Summary} closes the paper with a Summary.

\section{Year 1981: KORALB Monte Carlo for $e^+ e^- \to \tau^+\tau^- \gamma$ at Petra/PEP energies } \label{sec:KORALB}
At the time, Poland seemed to be an isolated place, but with enormous  in-flow of young talents  to
research. In reality a lot of contacts existed, but it was not to be seen by me.
Limited, and in fact quite awkward, access to computing existed. 
It looked like a hopeless loss of time, but a lot of
bright minds were attracted to the computing center of Jagellonian University.
I had access to algebraic manipulation language: {\tt Shoonship} \cite{Shoonship} too.

\begin{figure}[h]
\includegraphics[width=34pc]{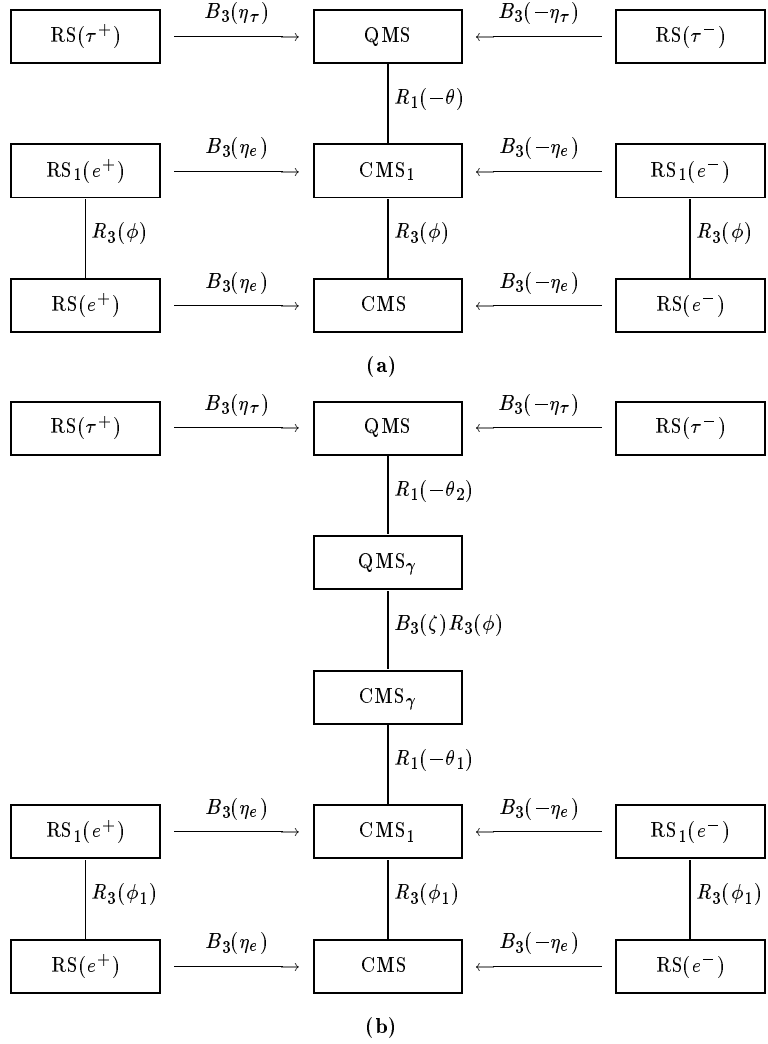}\hspace{2pc}%
\caption{\label{fig:KORALBframes} Tree of reference frames used to define amplitudes and Monte Carlo algorithm 
for $e^+ e^- \to \tau^+\tau^- \gamma$ process in refs.~\cite{Jadach:1984iy,Jadach:1985ac}.}
\end{figure}

One of my first projects was to evaluate the spin density matrix for the
process  $e^+ e^- \to \tau^+\tau^- \gamma$ at Petra/PEP energies \cite{Jadach:1984iy,Jadach:1985ac}
  This work was performed under the guidance of Prof. S. Jadach. 
This was quite an experience in looking at spin amplitudes as (reducible) 
representations of (Lorentz$\times$gauge) groups.
It was a great opportunity, programs were queuing for execution time. We could then
concentrate and  understand the details of what we were actually doing.
Later, in all my work, it came as an  enormous benefit:
how to represent  complicated formulas  (moderately complicated for today standards) 
of spin amplitudes  in a compact form. We have used a tree of reference frames to obtain the  goal, 
see Fig.~\ref{fig:KORALBframes}.  Instead of spin projections of 
individual fermions, we used differences (or sums) of such projections for 
incoming and/or outgoing leptons. In this way we could visualize spin 
properties of intermediate (formally virtual) photons.
Simplifications of lowest order amplitudes, 
remained  for bremsstrahlung as well. 
Nearly all calculations were done by hand, but every step was cross checked
with the help of algebraic calculation with  {\tt Shoonship}. This was an  
enormous help. We could also observe how some features of amplitudes could be identified if calculations 
were done by hand, but how easily these features were
overlooked if we relied too much on automatization.
Only if we knew what we were looking for, could we confirm the patterns with the algebraic tool. 
It is important to keep this in mind.
Similar problems are encountered in modern times with  applications of 
Machine Learning as well. I will return to this point in Section~\ref{sec:ML}.

\section{The t-channel contribution to spin amplitudes of s-channel Exclusive  Exponentiation and double gluon emission in QCD.} \label{sec:kkmc}

There is a multitude of factorization schemes available for  Field Theory calculations.
Over  years, I was checking, on particular examples, if hints for that could be 
identified already at the spin amplitude level. 
The general principle of the searches was rather simple. One tries 
to identify in amplitudes, gauge invariat   
 sub-structures (parts) responsible  for Matrix Element  enhancements: 
in some regions of the pase-space
(collinear-soft etc.). This is fundamental, especially from the point of view 
of Monte Carlo algorithm construction.
Discussions with {\bf Shimizu-sensei} were important for some stages of that work. 
The principle was to start from the complete expression, coded in a numerical program,  to identify 
the most singular term and then group some other terms necessary to obtain gauge invariant part of spin amplitude.
Then, for the remaining part, the search was repeated, until all  interesting parts  were 
identified. Starting point for that work had to be amplitudes guaranteed to be correct. 
In the two cases $e^+ e^- \to \nu_e \bar \nu_e \gamma \gamma$ \cite{Was:2004ig} and 
$q \bar q \to l^+ l^- g g$ \cite{vanHameren:2008dy}  expressions obtained with the help of algebraic programs 
were used for that purpose.

\begin{figure}[h]
\begin{minipage}[b]{24pc}
\includegraphics[width=20pc, angle=-90]{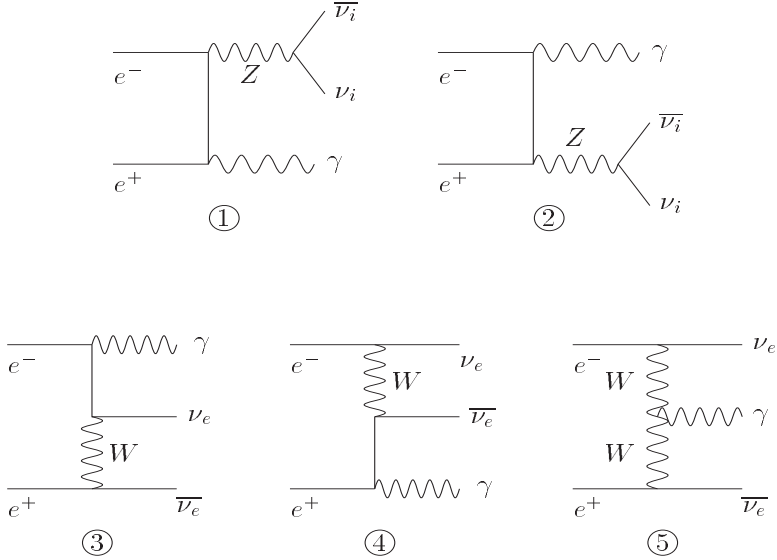}\hspace{2pc}%
\end{minipage}
\caption{ \label{fig:bremI}{ The Feynman diagrams for
          $e^+ e^- \rightarrow \bar \nu_e \nu_e \gamma$.}}
\end{figure}

Even though in principle there were no general rules to follow, separation of amplitudes into gauge invariant parts 
was a straightforward and seemingly unique procedure. However
I was not able to perform the task automatically. In fact, 
only some of the patterns appeared as a consequence of ordering singular terms. 
Feynman diagrams 1 and 2 of Fig.~\ref{fig:bremI}, combined, 
  complete the amplitude for 
 $e^+e^- \to \nu_\mu \bar \nu_\mu \gamma$ production, that is why they  form
gauge invariant part of amplitude for $e^+e^- \to \nu_e \bar \nu_e \gamma$ .


Further,    	
because these  two diagrams represent initial state QED bremsstrahlung
amplitude, they must   divide into parts, corresponding to $\beta_0, \beta_1$ of
Yennie-Frautshi-Suura exponentiation \cite{yfs:1961}. This has long been known, but the question was, whether it can be 
expanded to other cases, to higher orders or  to terms 
of different
singularities/enhancements than in case of QED photon emissions.
Unexpectedly, from  my experience 
 the answer seemed to be always ``yes''.
I could observe it not only in  QED and QCD cases 
but  also for example, in scalar QED \cite{Nanava:2006vv}. I was encouraged by Prof. Shimizu-sensei to follow this path, 
also in case of complete electroweak effects, and for loop contributions, but so far I have not found the solution.

Instead, let me  present first, single photon emission amplitude for 
$e^+e^- \to \nu_e \bar\nu_e \gamma$  (formula \ref{isr-feynman1}) 
and later for  double gluon emission in $q\bar q \to l^+l^-$ (formula \ref{eq:Mab}). 
 For notation conventions references
\cite{Was:2004ig,vanHameren:2008dy} are used.  

\begin{eqnarray}
  \label{isr-feynman1}
  && \Meu_{1\{I\}}\left( \st^{p}_{\lambda} \st^{k_1}_{\sigma_1} \right)=
    {\cal M}^0+{\cal M}^1+{\cal M}^2+{\cal M}^3,
   \\
    {\cal M}^0&=&
   eQ_e \;
    \bar{v}(p_b,\lambda_b)\; \mathbf{M}^{bd}_{\{I\}}\;
         {\not\!{p_a}+m-{\not\!k_1} \over -2k_1p_a} \not\!{\epsilon}^\star_{\sigma_1}(k_1)\;
    u(p_a,\lambda_a)
    \nonumber \\&&
   +eQ_e \;
    \bar{v}(p_b,\lambda_b)
         \not\!{\epsilon}^\star_{\sigma_1}(k_1)\; {-{\not\!p_b}+m+\not\!{k_1} \over -2k_1p_b} \;
         \mathbf{M}^{ac}_{\{I\}}\;
    u(p_a,\lambda_a),
    \nonumber\\
       {\cal M}^{1}&=&{\cal M}^{1'}+{\cal M}^{1''},  \nonumber  \\
       {\cal M}^{1'}&=&+e \;
    \bar{v}(p_b,\lambda_b)
        \; \;
         \mathbf{M}^{bd,ac}_{\{I\}}\;
    u(p_a,\lambda_a) {\epsilon}^\star_{\sigma_1}(k_1) \cdot (p_c-p_a) {1 \over {t_a -M_W^2}}{1 \over {t_b -M_W^2}},
    \nonumber \\
       {\cal M}^{1''}&=&+e \;
    \bar{v}(p_b,\lambda_b)
        \; \;
         \mathbf{M}^{bd,ac}_{\{I\}}\;
    u(p_a,\lambda_a) {\epsilon}^\star_{\sigma_1}(k_1) \cdot (p_b-p_d) {1 \over {t_a -M_W^2}}{1 \over {t_b -M_W^2}},
    \nonumber\\
    {\cal M}^2&=&   +e \; \bar{v}(p_b,\lambda_b)  g_{\lambda_b,\lambda_d}^{We\nu}
         \not\!{\epsilon}^\star_{\sigma_1}(k_1)\;  \;
    v(p_d,\lambda_d)
    \bar{u}(p_c,\lambda_c)  g_{\lambda_c,\lambda_a}^{We\nu}
         \not\!k_1\;  \;
    u(p_a,\lambda_a) {1 \over {t_a -M_W^2}}{1 \over {t_b -M_W^2}},
   \nonumber  \\
     {\cal M}^3&=&  -e \; \bar{v}(p_b,\lambda_b)  g_{\lambda_b,\lambda_d}^{We\nu}
         \not\!k_1\;  \;
    v(p_d,\lambda_d)
    \bar{u}(p_c,\lambda_c) g_{\lambda_c,\lambda_a}^{We\nu}
         \not\!{\epsilon}^\star_{\sigma_1}(k_1)\;  \;
    u(p_a,\lambda_a) {1 \over {t_a -M_W^2}}{1 \over {t_b -M_W^2}},\nonumber 
 \end{eqnarray}
Once manipulations are completed, we separate the  spin amplitude 
 into {\it six} individually QED gauge invariant parts. This is rather
easy to check, replacing photon polarization vector with its 
four-momentum. Each of the obtained parts  has a well defined physical 
interpretation. 
It is also possible to verify that the  gauge invariance of each part
can be preserved in  the case of the extrapolation, necessary in case of QED exclusive exponentiation
(see Refs.~\cite{Jadach:1999vf,Jadach:2000ir}). Then, because of 
additional photons, the condition $p_a+p_b=p_c+p_d+k_1$ is not valid.

Let us now turn to double gluon emission amplitude, that is for the process $q\bar q \to l^+l^- g g$:

\begin{equation} \label{eq:Mab}
\Mab
=
\frac{1}{2}
\,\bar{v}(\pp)
\Big(
\Ta\Tb \IAB +\Tb\Ta \IBA
\Big)
u(\qq)
~.
\end{equation}
For the $\Ta\Tb$-part, we find
\begin{eqnarray}
\IAB
&=&
\bigg(
 \frac{\inp{\pp}{\eA}}{\inp{\pp}{\kA}}
-\frac{\inp{\kB}{\eA}}{\inp{\kB}{\kA}}
-\frac{\esA\ksA}{2\inp{\pp}{\kA}}
\bigg)
\Jsl
\bigg(
 \frac{\ksB\esB}{2\inp{\qq}{\kB}}
+\frac{\inp{\kA}{\eB}}{\inp{\kA}{\kB}}
-\frac{\inp{\qq}{\eB}}{\inp{\qq}{\kB}}
\bigg)
\nonumber 
\\
&+&
\frac{\inp{\pp}{\kB}}{\inp{\pp}{\kA}+\inp{\pp}{\kB}-\inp{\kA}{\kB}}
\,\bigg(\frac{\inp{\pp}{\eA}}{\inp{\pp}{\kA}}-\frac{\inp{\kB}{\eA}}{\inp{\kB}{\kA}}
        -\frac{\esA\ksA}{2\inp{\pp}{\kA}}\bigg)
  \bigg(\frac{\inp{\pp}{\eB}}{\inp{\pp}{\kB}}-\frac{\inp{\kA}{\eB}}{\inp{\kA}{\kB}}
        -\frac{\esB\ksB}{2\inp{\pp}{\kB}}\bigg)
\Jsl
\nonumber 
\\
&+&
\Jsl 
\,\frac{\inp{\qq}{\kA}}{\inp{\qq}{\kA}+\inp{\qq}{\kB}-\inp{\kA}{\kB}}
\,\bigg(\frac{\inp{\qq}{\eA}}{\inp{\qq}{\kA}}-\frac{\inp{\kB}{\eA}}{\inp{\kB}{\kA}}
        -\frac{\ksA\esA}{2\inp{\qq}{\kA}}\bigg)
  \bigg(\frac{\inp{\qq}{\eB}}{\inp{\qq}{\kB}}-\frac{\inp{\kA}{\eB}}{\inp{\kA}{\kB}}
        -\frac{\ksB\esB}{2\inp{\qq}{\kB}}\bigg)
\nonumber 
\\
&+&
\Jsl 
\bigg(
1
 -\frac{\inp{\pp}{\kB}}{\inp{\pp}{\kA}+\inp{\pp}{\kB}-\inp{\kA}{\kB}}
-\frac{\inp{\qq}{\kA}}{\inp{\qq}{\kA}+\inp{\qq}{\kB}-\inp{\kA}{\kB}}
\bigg)
\bigg( \frac{\inp{\kA}{\eB}}{\inp{\kA}{\kB}}\,\frac{\inp{\kB}{\eA}}{\inp{\kA}{\kB}}
      -\frac{\inp{\eA}{\eB}}{\inp{\kA}{\kB}}\bigg)
\nonumber 
\\
&-&
\frac{1}{4}
\,\frac{1}{\inp{\pp}{\kA}+\inp{\pp}{\kB}-\inp{\kA}{\kB}}
\bigg(
  \frac{ \esA\ksA\esB\ksB
         -\esB\ksB\esA\ksA }{\inp{\kA}{\kB}}
\bigg)
\Jsl
\nonumber 
\\
&-&
\frac{1}{4}
\,\Jsl 
\,\frac{1}{\inp{\qq}{\kA}+\inp{\qq}{\kB}-\inp{\kA}{\kB}}
\bigg(
  \frac{ \ksA\esA\ksB\esB
         -\ksB\esB\ksA\esA }{\inp{\kA}{\kB}}
\bigg)
\label{abend}
~.
\end{eqnarray}
%
The part proportional to other order of SU(3) group generators \formula{$\Tb\Ta$}, is obtained by a permutation of 
the momenta and polarization vectors of the gluons. 
Each line of the above expression is individually gauge invariant. Also, this expression for double
gluon emission is rather compact too.

\section{The $e^+ e^- \to 4f$ process}  \label{sec:4f}

\begin{figure}[h]
\begin{minipage}{20pc}
\includegraphics[width=20pc, angle=-90]{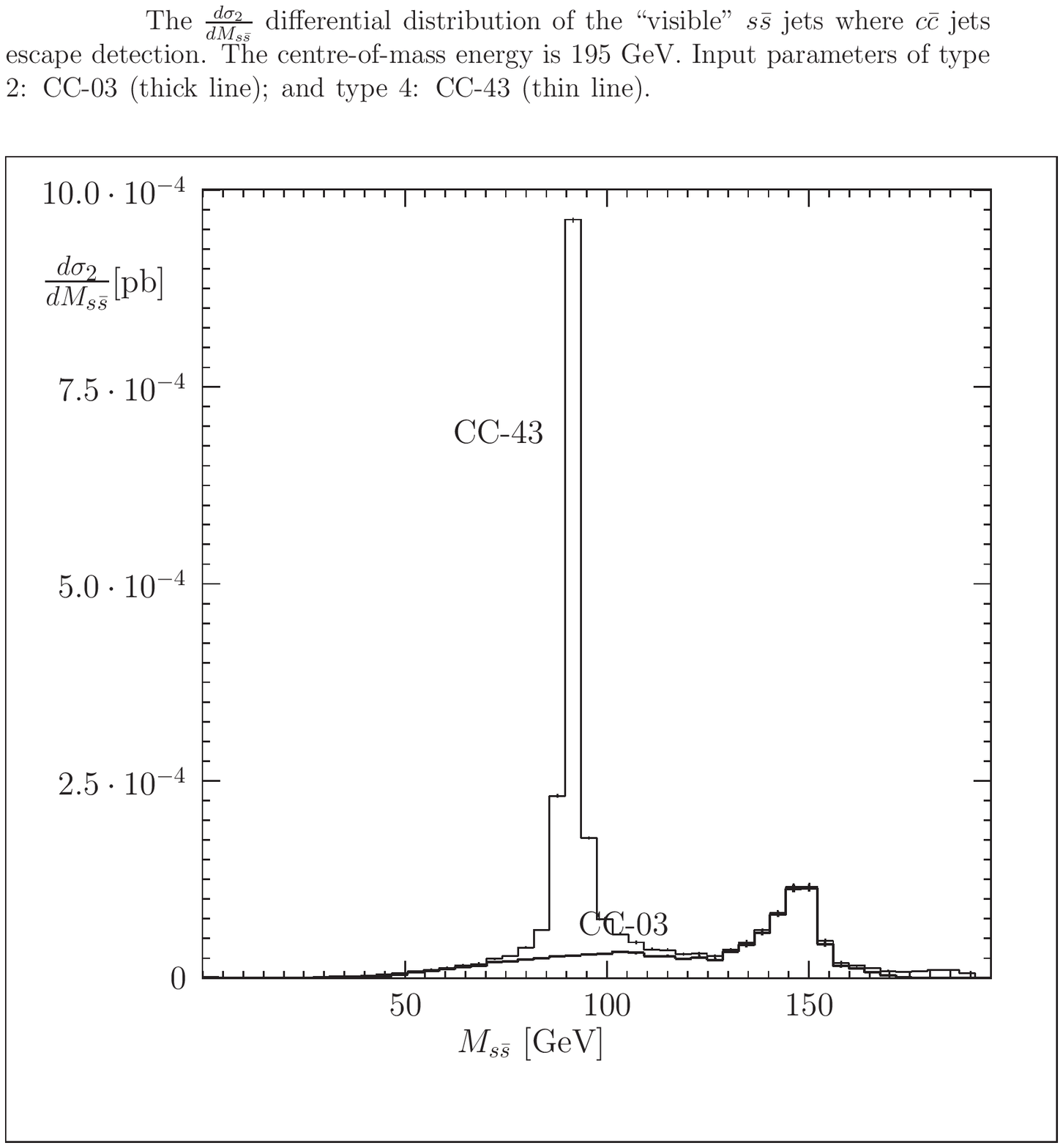}
\end{minipage}
\begin{minipage}{20pc}
\includegraphics[width=20pc, angle=-90]{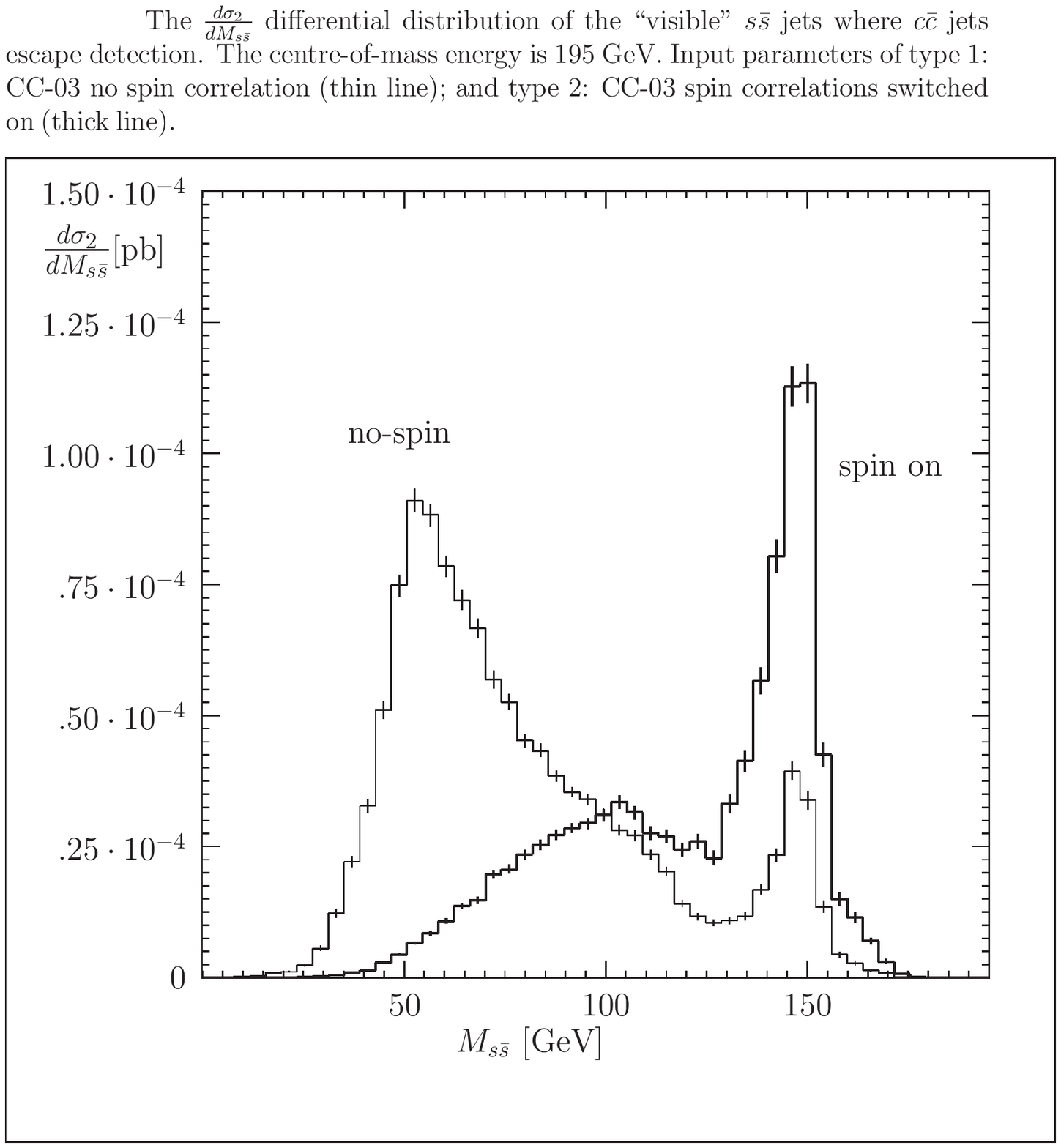}
\end{minipage}
\caption{ \label{fig:ww} Figure taken from Ref.~\cite{Ishikawa:1997ma}}
\end{figure}

The main purpose of my  visit  to  KEK  MinamiTateya group in 1996, was to work 
on  {\tt Grace} spin amplitudes \cite{Fujimoto:2002sj} and to prepare them for use 
in our {\tt KORALW} Monte Carlo \cite{Jadach:2001mp} for the
$ e^+e^- \to 4$~fermion processes at LEP II energies.
In this work, because of Monte Carlo integration, phase space regions of collinear configurations, 
resulted in numerical difficulties. 
This required careful and painful work to avoid `trivial' mistakes due to rounding errors damaging gauge 
cancellations. On the margins of this work,
kinds of fake `New Physics', phenomena appeared.
Let me show, for myself  at first rather unexpected example, see Ref.~\cite{Ishikawa:1997ma}.
It illustrates that the interplay of theoretical effects and selection cuts can be confusing. In 
$e^+e^- \to WW \to q \bar q q \bar q $ two jets were requested to be lost in the beam pipe the 
invariant mass of the other two was monitored. If all diagrams of the lowest order Standard Model amplitudes 
were taken into account, a clear $Z$ peak is present (Fig.~\ref{fig:ww} left side). But there is another peak at a much 
higher mass, which  
do not disappear, even if only $W$-pair production and decay amplitudes are taken into account
(see Fig.~\ref{fig:ww} right side). It is a consequence of the veto cut, $W^+$
and $W^-$ peaks, and spin correlations  
as explained in Ref. \cite{Ishikawa:1997ma}.

Such phenomena need to be included in generators used for simulation of backgrounds. Otherwise intuition may 
fail to recognize importance for background estimation effects such as generally small spin correlations.
Cross-check with {\tt Grace} spin amplitudes was helpful to confirm the origin of the phenomenon.

\section{Machine Learning for Higgs parity measurement in $H \to \tau \tau$ decay}  \label{sec:ML}

Let us now turn to  another example, where complex observables
need to be defined and computer techniques of the Neural Network or Machine Learning (ML) 
type \cite{lecun2015deep,kingma2014adam,ioffe2015batch,srivastava2014dropout,abadi2015tensorflow}
are useful or even irreplaceable. Fifteen years ago, in Ref.~\cite{Bower:2002zx}, we have proposed to measure
Higgs boson parity in its $H \to \tau \tau$ decay with the help of acoplanarity angle for planes built on visible 
decay products for consecutive $\tau^\pm \to \pi^\pm \pi^0 \nu_\tau$ decays. 
A variant of the observable where impact parameter was used, is presented in \cite{Desch:2003mw}.  
In  attempt to extend the method to
$\tau\to 3\pi \nu$ decays, one can construct  4 or 16 such angles for each event.
Each of these distributions provide some CP sensitivity,
but distributions are  correlated and  possible backgrounds will complicate future measurements
even further.
Such observables of multidimensional nature may  be  
controlled today with ML techniques. An attempt in this direction was presented 
in Ref.~\cite{Jozefowicz:2016kvz}. In the following, let us recall some 
details of this work.

The $H$ or $A$ parity information, thanks to the sign difference, can be extracted from  
the correlations between $\tau^{+}$ and $\tau^{-}$ transverse spin components. 
The decay probability 
\[
\Gamma(H/A\to \tau^{+}\tau^{-}) 
\sim 1-s^{\tau^{+}}_{\parallel}  
s^{\tau^{-}}_{\parallel}\pm s^{\tau^{+}}_{\perp}s^{\tau^{-}}_{\perp}
\]
is sensitive to the $\tau^\pm$ polarization   
vectors $s^{\tau^{-}}$ and $s^{\tau^{+}}$  (defined    
in their respective rest frames).    
The symbols {\scriptsize ${\parallel}$,${\perp}$} denote components
 parallel/transverse   
to the Higgs boson momentum as seen from the respective $\tau^\pm$  
rest frames.

Because of the narrow $\tau$ width,
cross-section for the process $ f \bar f \to \tau^+\tau^-  Y ; 
                              \tau^+ \to X^+ \bar \nu; \tau^-\to X^- \nu$ reads:
 \[ 
d \sigma = \sum_{spin }|{\cal M}|^2 d\Omega= \sum_{spin }|{\cal M}|^2 d\Omega_{prod} \; d\Omega_{\tau^+} \; d\Omega_{\tau^-}.
\]
With  only $\tau$ spin indices explicit $\cal M$ reads:
\[ 
{\cal M}=\sum_{\lambda_1\lambda_2=1}^2{\cal M}_{\lambda_1\lambda_2}^{prod} \; 
      {\cal M}_{\lambda_1}^{\tau^+}{\cal M}_{\lambda_2}^{\tau^-}.
\]
 The expression for $d \sigma$ can be re-written into core formula of spin algorithms of $\tau$-pair production and decay:
 \[ 
d \sigma = \Bigl(\sum_{spin }|{\cal M}^{prod}|^2 \Bigr)
 \Bigl(\sum_{spin }|{\cal M}^{\tau^+}|^2 \Bigr)
 \Bigl(\sum_{spin }|{\cal M}^{\tau^-}|^2 \Bigr) wt \;
 d\Omega_{prod} \; d\Omega_{\tau^+} \; d\Omega_{\tau^-}
\]

To complete explanations, we need to first
recall, details of general formalism of semileptonic $\tau$-lepton
decays. The Matrix Element used in {\tt TAUOLA} Monte Carlo \cite{Jadach:1993hs}
for semileptonic decay
$\tau(P,s)\rightarrow\nu_{\tau}(N)X$  can be written as follows:
${\cal M}=\frac{G}{\sqrt{2}}\bar{u}(N)\gamma^{\mu}(v+a\gamma_5)u(P)J_{\mu}$, 
the current $J_{\mu}$ depends on the momenta of all hadrons. Then
\begin{eqnarray}
 |{\cal M}|^{2}&=& G^{2}\frac{v^{2}+a^{2}}{2}( \omega + H_{\mu}s^{\mu} ), \\
 \omega&=&P^{\mu}(\Pi_{\mu}-\gamma_{va}\Pi_{\mu}^{5}), \nonumber \\
 H_{\mu}&=&\frac{1}{M}(M^{2}\delta^{\nu}_{\mu}-P_{\mu}P^{\nu})(\Pi_{\nu}^{5}-
\gamma_{va}\Pi_{\nu}),\nonumber \\
 \Pi_{\mu}&=&2[(J^{*}\cdot N)J_{\mu}+(J\cdot N)J_{\mu}^{*}-(J^{*}\cdot J)
N_{\mu}],\nonumber \\
 \Pi^{5\mu}&=&2~ {\rm Im} ~\epsilon^{\mu\nu\rho\sigma}J^{*}_{\nu}J_{\rho}N_{\sigma}, \nonumber\\
\gamma_{va}&=&-\frac{2va}{v^{2}+a^{2}},\nonumber\\
\hat{\omega}&=&2\frac{v^{2}-a^{2}}{v^{2}+a^{2}}m_{\nu}M(J^{*} \cdot J),\nonumber\\
\hat{H}^{\mu}&=&-2\frac{v^{2}-a^{2}}{v^{2}+a^{2}}m_{\nu}~ {\rm Im}~\epsilon^{\mu\nu\rho\sigma}
J_{\nu}^{*}J_{\rho}P_{\sigma}.\nonumber
\end{eqnarray}
In the following we will use $h_i=H^i/H_0$.

The Higgs decay probability in the formalism of Refs.~\cite{Kramer:1993jn,Desch:2003rw} in case
when both scalar and pseudo-scalar $H\tau\tau$ coupling are allowed
$\bar{\tau}N(\cos\phi^{CP}+i\sin\phi^{CP}\gamma_{5})\tau$,
reads
\begin{equation}
\Gamma(h_{mix}\to \tau^{+}\tau^{-}) \sim 1-s^{\tau^{+}}_{\parallel}
s^{\tau^{-}}_{\parallel}+ s^{\tau^{+}}_{\perp} 
R(2\phi^{CP})~s^{\tau^{-}}_{\perp}.
\end{equation}
The  $R(2\phi^{CP})$ $-$ denotes the operator for the rotation by
angle $2\phi^{CP}$  around the {\footnotesize ${\parallel}$} direction.

As a consequence, spin weight $wt$ for the combined production and decay of tau pair reads as:
\begin{equation}
wt= 1-h^{\tau^{+}}_{\parallel}
h^{\tau^{-}}_{\parallel}+ h^{\tau^{+}}_{\perp} 
R(2\phi^{CP})~h^{\tau^{-}}_{\perp}.
\end{equation}
Naturally, the Higgs parity  should  
reflects itself in some kind of correlations between    
the $\tau^+$ and $\tau^-$ decay products in the directions transverse to the   
$\tau^{+}\tau^{-}$ axes.

\begin{figure}[h]
\begin{minipage}{18pc}
\includegraphics[width=18pc, angle=-90]{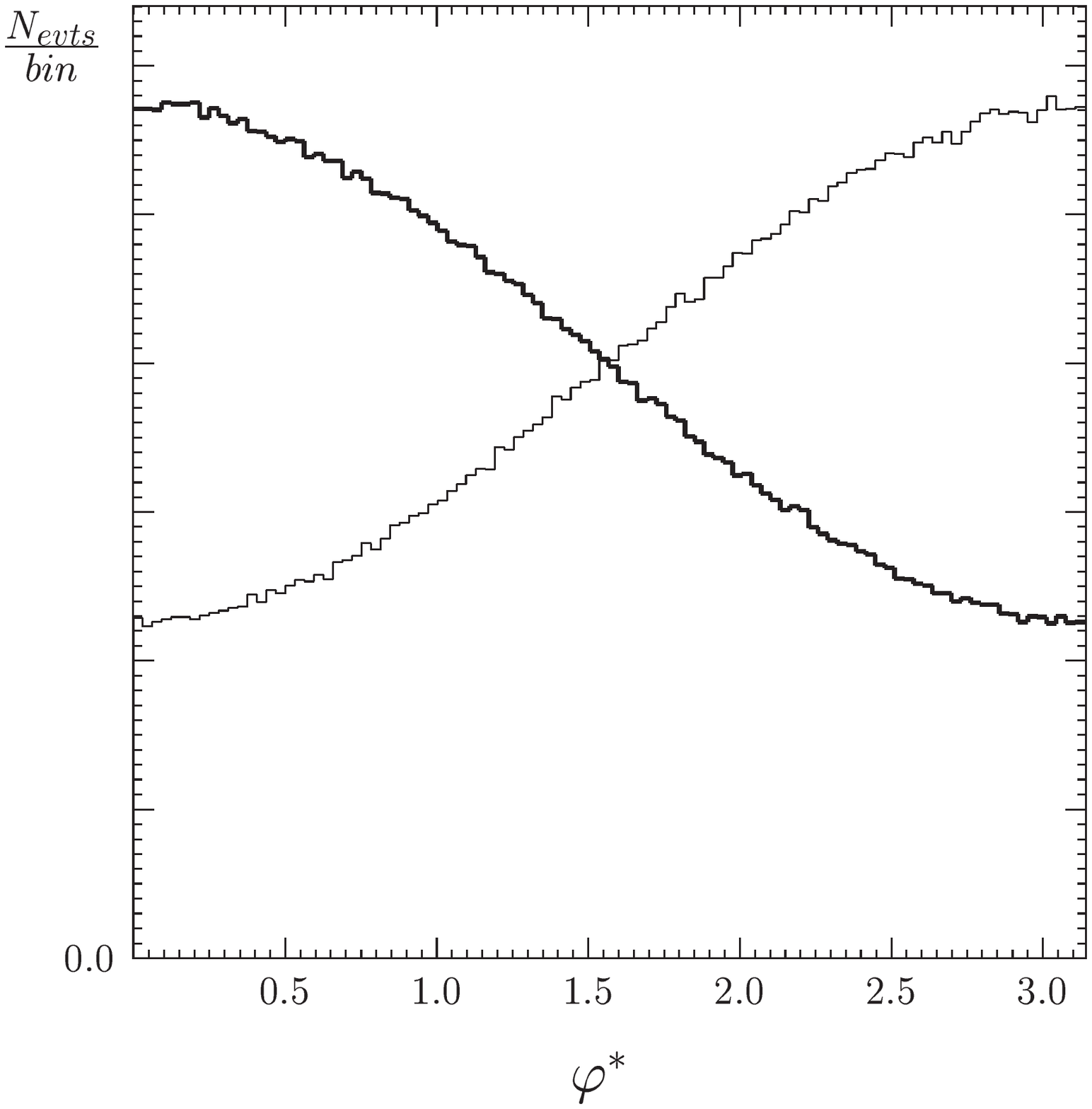}
\centerline{ $y_1*y_2>0$ }
\end{minipage}
\begin{minipage}{18pc}
\includegraphics[width=18pc, angle=-90]{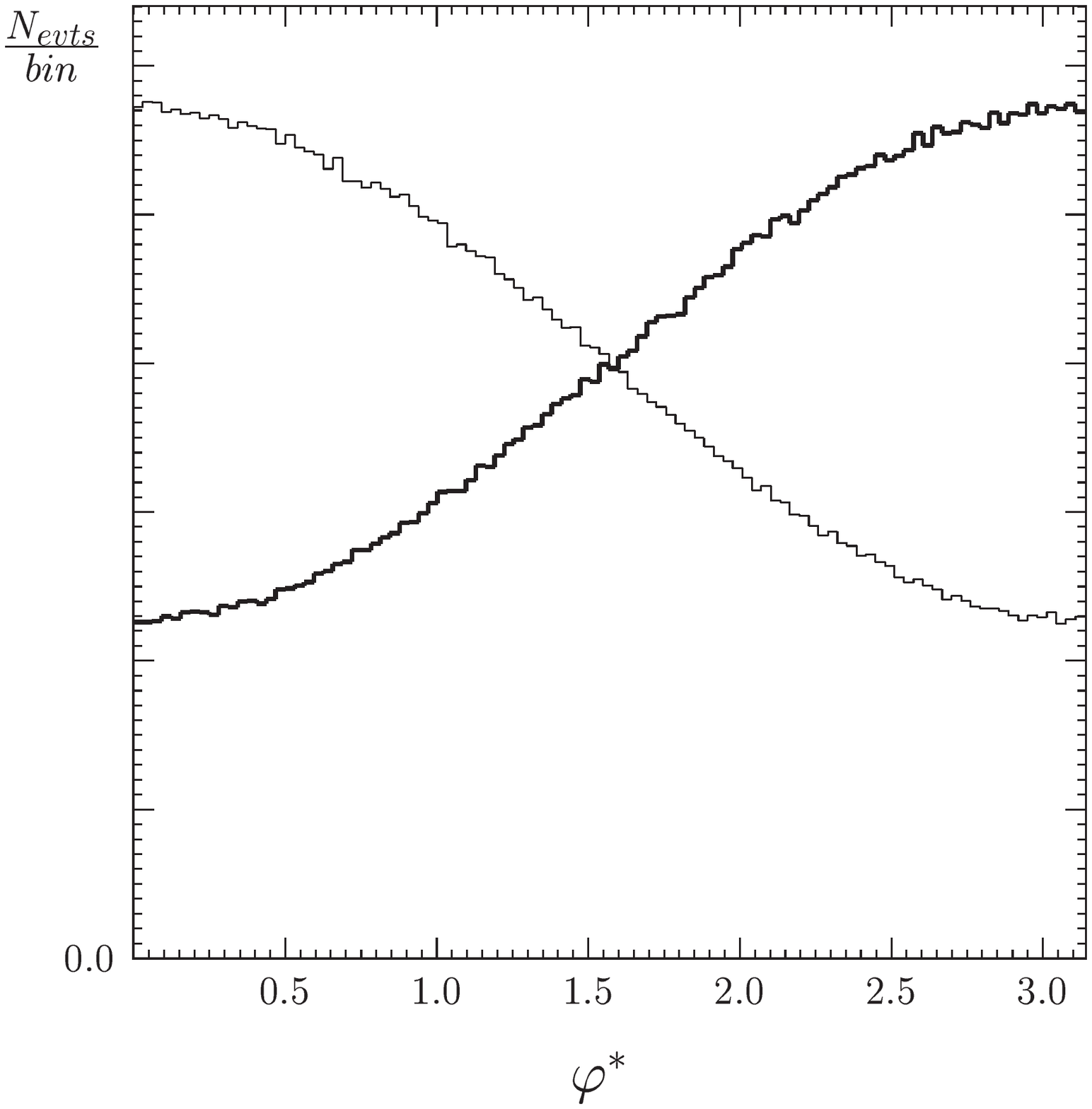}
\centerline{$y_1*y_2 <0$}
\end{minipage} 
\caption{\label{fig:rho-rho}The ~$\rho^+ \rho^-$ decay products' acoplanarity distribution
in $\rho^+ \rho^-$ pair rest-frame. Thick line denote the case   
of the scalar Higgs  and thin line the pseudo-scalar.}
\end{figure}

\begin{figure}[h]
\includegraphics[width=18pc]{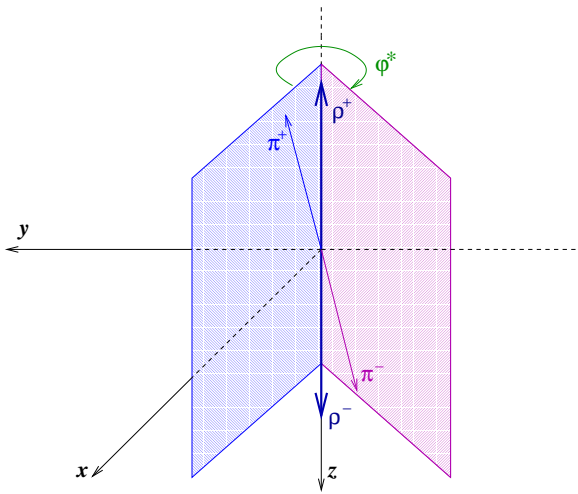}\hspace{2pc}%
\begin{minipage}[b]{10pc}\caption{\label{fig:acopl} Acoplanarity angle
 $0<\varphi^*<2\pi$ between oriented half-planes (or $0<\varphi^*<\pi$ between oriented planes) spanned respectively 
on $\rho^\pm-\pi^\pm$ decay products and in the rest frame of $\rho^+\rho^-$ pair.}
\end{minipage}
\end{figure}

Let us recall first the case when both $\tau$ leptons decay to $\pi^\pm\pi^0 \nu_\tau$.
Then $h^i =  {\cal N} \Bigl( 2(q\cdot N)  q^i -q^2  N^i \Bigr)   $, where 
$q\cdot N = (E_{\pi^\pm} - E_{\pi^0}) m_\tau$ and four momentum  
$q=p_{\pi^\pm} - p_{\pi^0}$ is build from four momenta of $\pi$'s. The  $N$ denote 
four momentum of neutrino again in the $\tau$ lepton rest-frame. 
Because  $q\cdot N$ may be either positive or 
negative,  corresponding 
regions of phase space would contribute cancelling out parity effects. To separate 
we may use   $y_1$, $y_2$ variables. The
$y_1={E_{\pi^{+}}-E_{\pi^{0}}\over E_{\pi^{+}}+E_{\pi^{0}}}~;~ 
y_2={E_{\pi^{-}}-E_{\pi^{0}}\over E_{\pi^{-}}+E_{\pi^{0}}}$ can be calculated directly 
from $\pi$'s energies as measured in the laboratory frame.
It is enough to take the 
sign of the $y_1 y_2$ only. The sensitivity
to parity can be seen in  Fig.~\ref{fig:rho-rho}. Acoplanarity
angle (see Fig.~\ref{fig:acopl} for definition) was used. Already in this case,
the observable is in principle of 3-dimensional nature.

\begin{figure}[h]
\begin{minipage}{12pc}
\includegraphics[width=12pc]{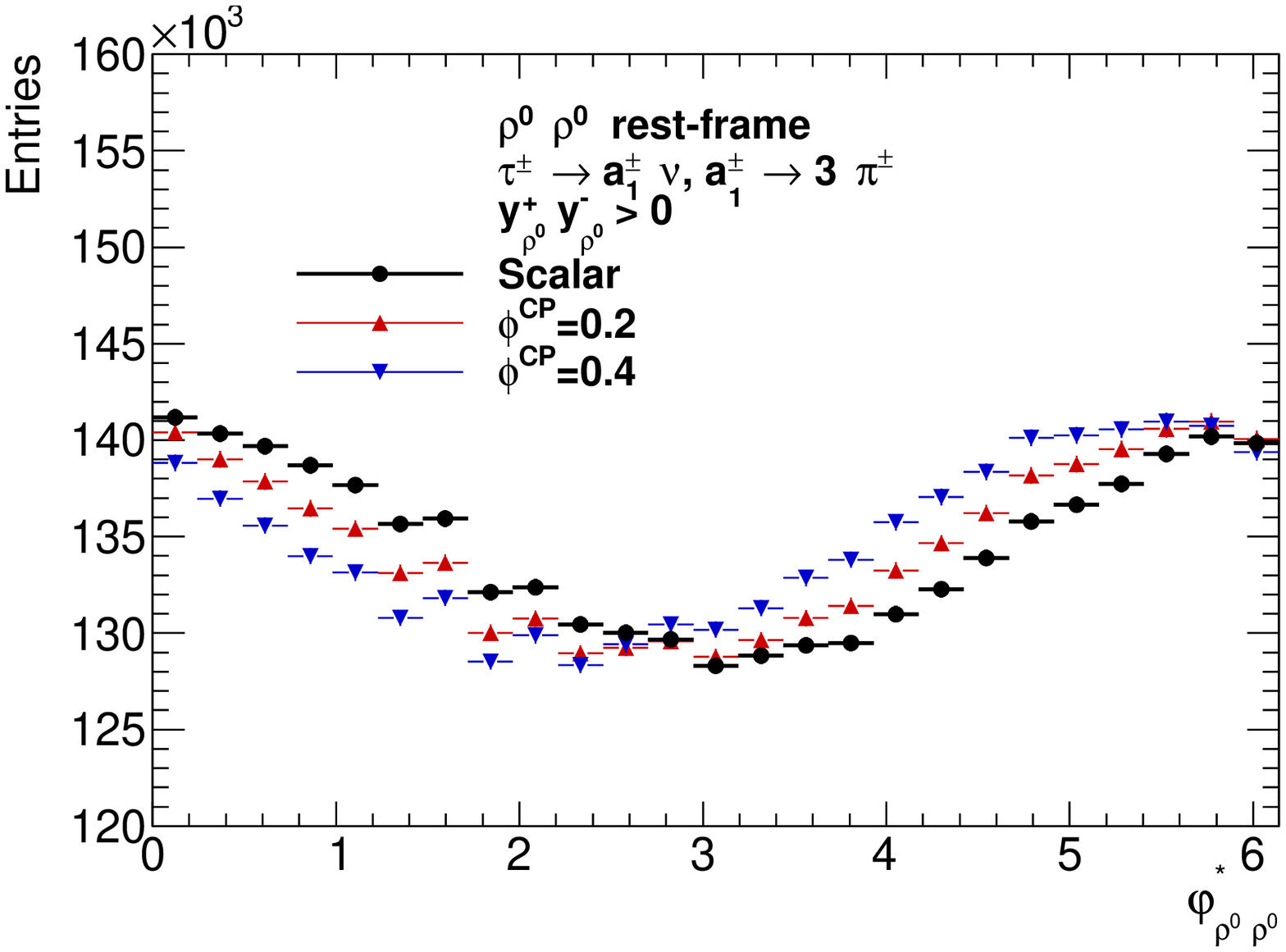}
\end{minipage}
\begin{minipage}{12pc}
\includegraphics[width=12pc]{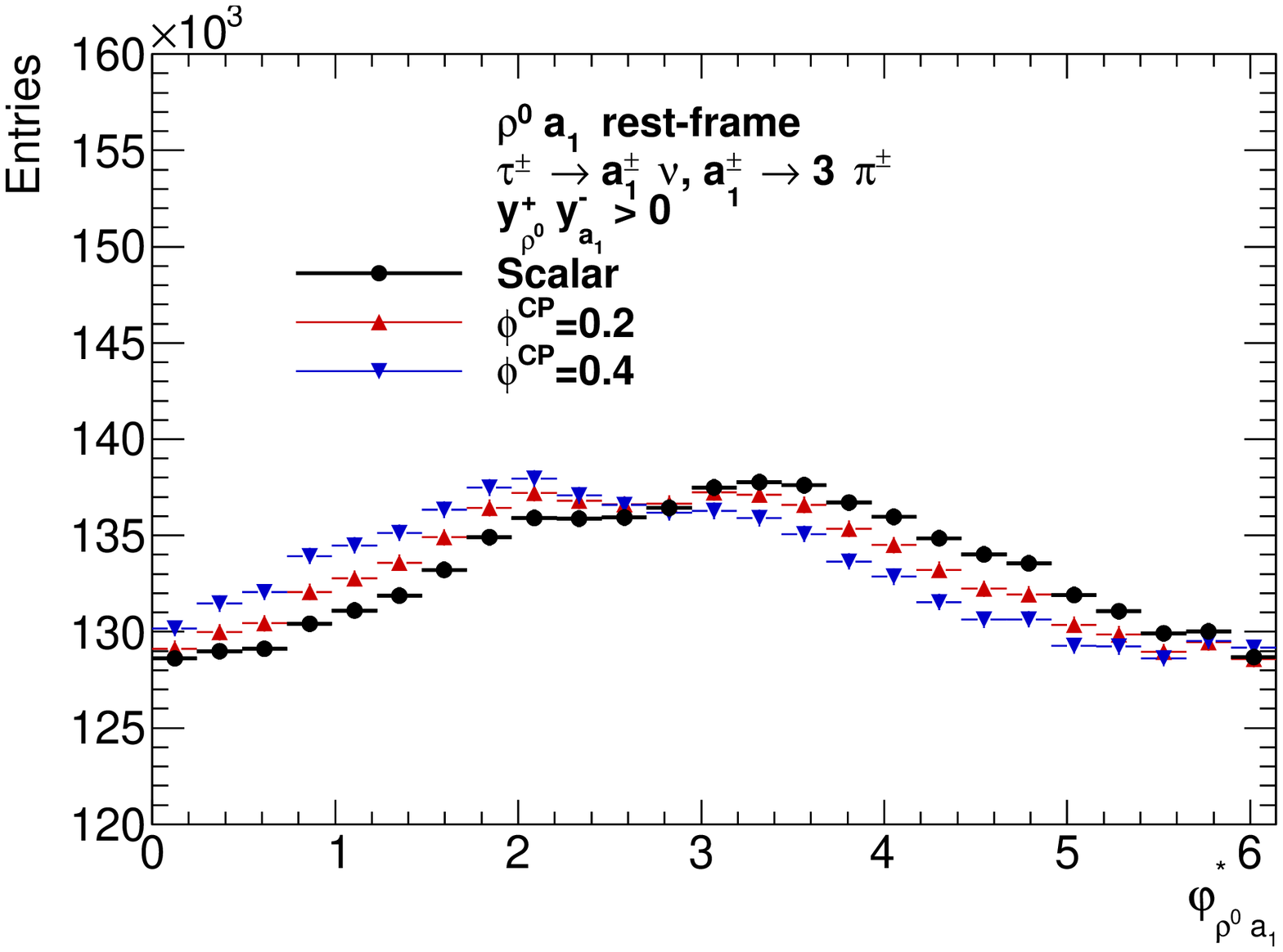}
\end{minipage} 
\begin{minipage}{12pc}
\includegraphics[width=12pc]{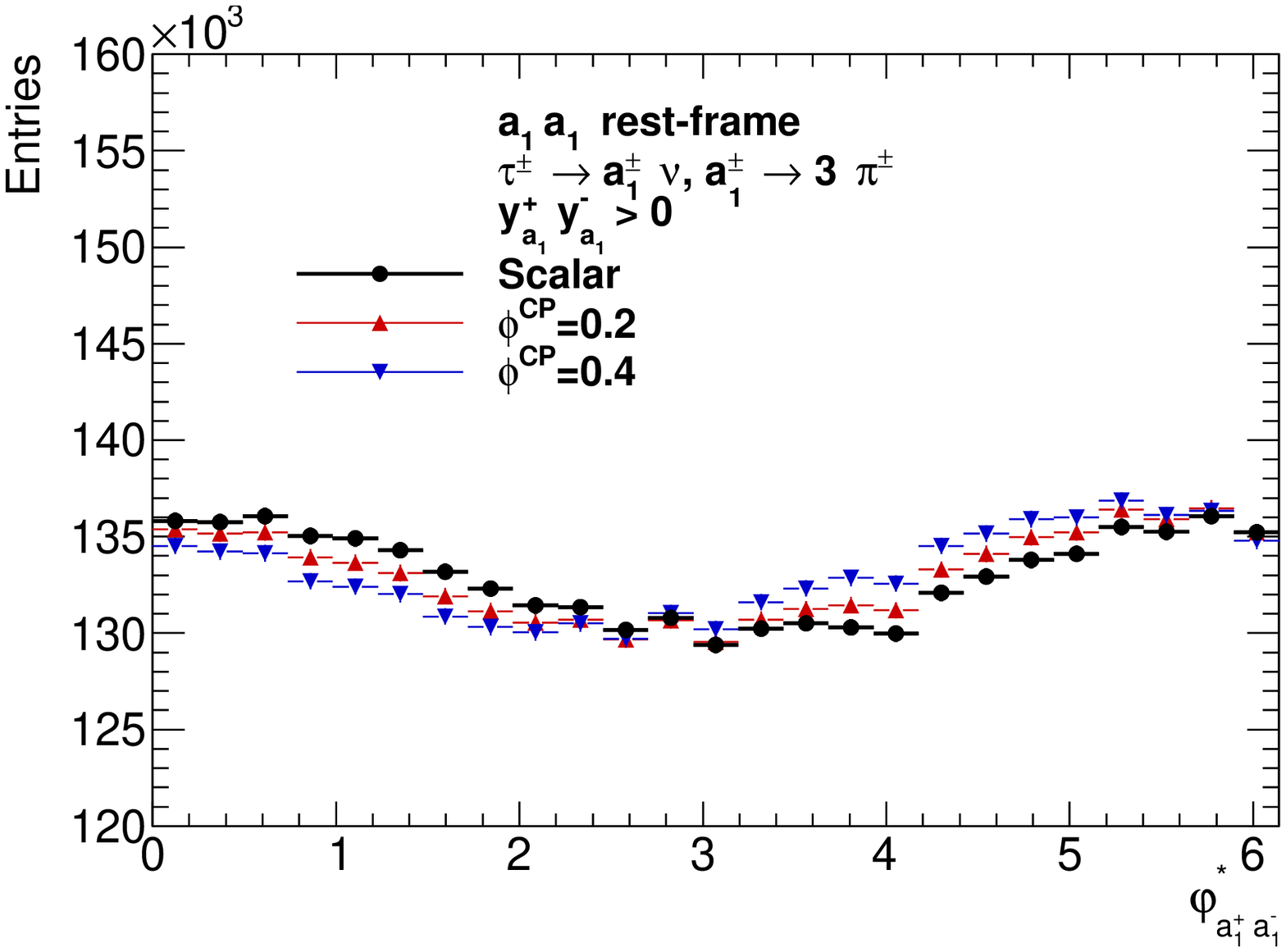}
\end{minipage} 
\caption{\label{fig:a1-a1}Acoplanarity angles of oriented half decay planes: $\varphi^*_{\rho^{0} \rho^{0}}$ (left), $\varphi^*_{a_1 \rho^{0}}$ (middle) and  $\varphi^*_{a_1 a_1}$ (right), 
for events grouped by the sign of  $y_{\rho^{0}}^+ y_{\rho^{0}}^-$, $y_{a_1}^{+} y_{\rho^{0}}^-$ and $y_{a_1}^{+} y_{a_1}^{-}$ respectively. Three CP mixing angles  $\phi^{CP}$ = 0.0 (scalar), 0.2 and 0.4. 
Note that physics model depends on 1 parameter only and effect of  $\phi^{CP}$,
the Higgs mixing scalar pseudo-scalar angle,  is  a linear shift. }
\end{figure}

\begin{figure}[h]
\vskip -2 cm
\includegraphics[width=24pc]{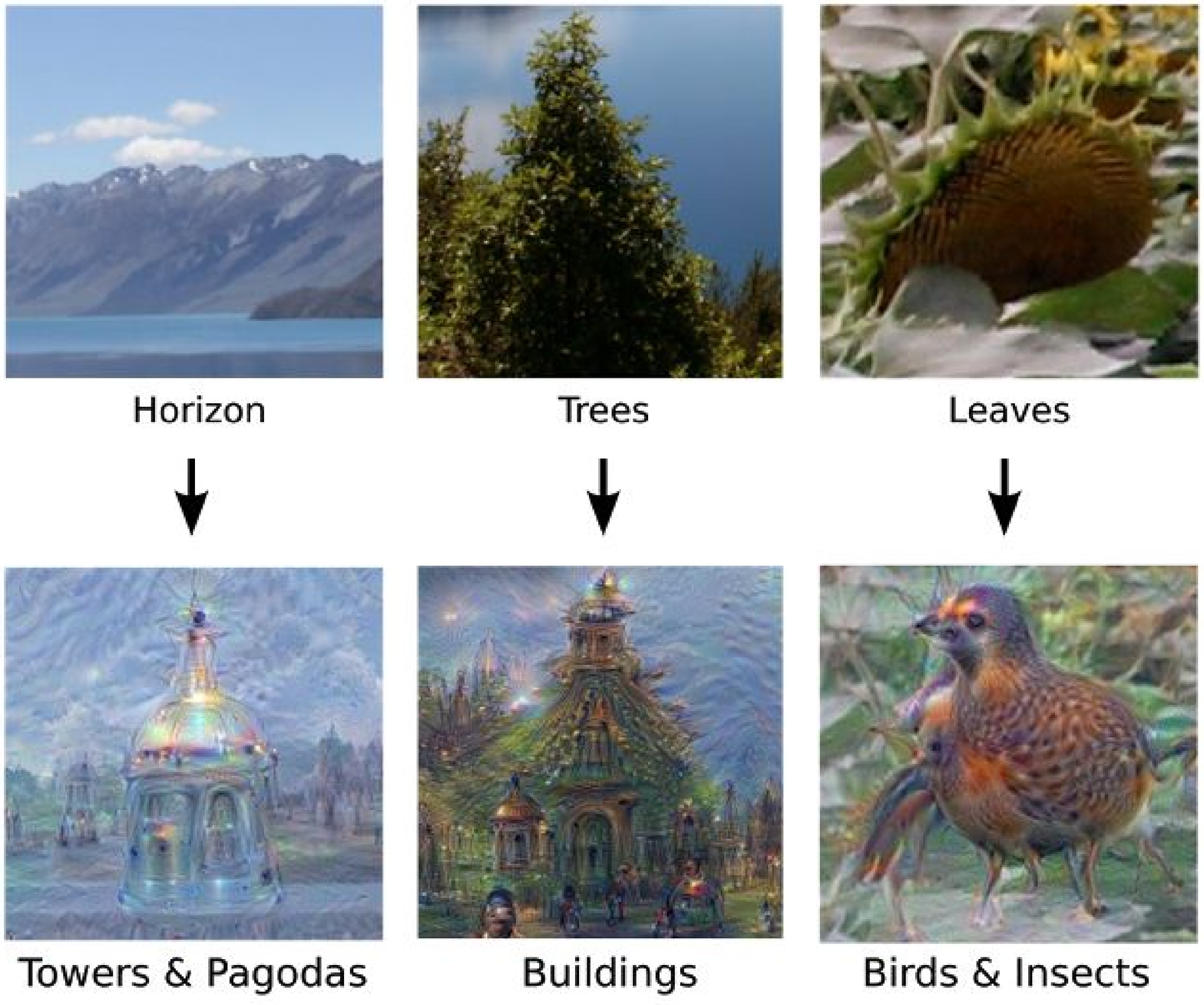}\hspace{2pc}%
\caption{ \label{fig:ANN} \scriptsize Artificial Neural Networks have spurred remarkable recent progress in image 
classification and speech recognition. But even though these are very useful 
tools based on well-known mathematical methods, we actually understand 
surprisingly little of why certain models work and others don't. 
\newline
{ From 
http://googleresearch.blogspot.com/2015/06/inceptionism-going-deeper-into-neural.html }
\newline {  Pattern recognition is an active field
and deep concern and not only for us.}}
\end{figure}

In case of $\tau\to 3 \pi\nu $ decay products, {\it four} distinct planes can be spanned on its visible decay products, 
thus if both $\tau^+$ and $\tau^-$ decay in this chanel,   16 acoplanarity angles can be defined, see Fig.~\ref{fig:a1-a1}, and also 8 variables similar to
$y_{1,2}$. The observable is thus built on 24 dimensional space. 
Fortunately, with the modern techniques, such as used in 
Ref.~\cite{Jozefowicz:2016kvz} an overall sensitivity can be evaluated. On the other hand,
as it may be in general difficult to develop intuition, the risk of misinterpretation can not be ignored. 
Difficulties of all pattern recognition projects are well known, see Fig.~\ref{fig:ANN}. These 
challenges, were explored already by  Giuseppe Arcimboldo (1572 - 1593) in his paintings.

\section{Summary}  \label{sec:Summary}

Inspired by Shimizu-sensei conference, I have focused on  an aspect of work for  Monte Carlo 
generators and on phenomenology of High Energy Physics experiments related to the use  
of algebraic manipulation programs. 
 My aim was  to show several simple examples of challenges, resulting 
from complexity: how automated calculations were of help, but also sources 
of difficulties.
Some of the important  examples  originate from my work in Minami Tateya group 
I was visiting over the last 25 years.
Each example deserve substantial introduction. This was impossible for a short talk.
Whenever possible, I delegate the reader  to references.
Inescapably, I have presented only scattered projects,  where use of computer algebraic 
methods or  pattern recognition techniques (Machine Learning) were necessary.
Collecting material  for  my talk was inspiring and educative for myself as  I could look 
back at particular
context of my scientific activities over all these years.


\ack
{\small
This talk would not be possible without contribution and collaboration with the co-authors of the presented papers: S. Jadach, E. Richter-Was, M. Skrzypek
and  B. F. L. Ward.
I would like to stress importance of research which was performed in collaboration 
with T. Ishikawa and  Y. Kurihara at times when I was visiting Minanmi Tateya group of Prof. Y. Shimizu. This collaboration  was essential part of my experience
with algebraic manipulation programs and programs constructed with their help.

Work supported in part by funds of Polish National Science
Centre under decisions UMO-2014/15/ST2/00049 and by the Research Executive 
Agency (REA) of the European Union under the Grant Agreement PITNGA2012316704 (HiggsTools).
I would like to express gratitude to  the conference organizers for the invitation.
}

\section*{References}
\providecommand{\newblock}{}

\end{document}